\documentclass[12pt,preprint]{emulateapj}
\usepackage{natbib,subfigure}
\newcommand{\myemail}{p.cargile@vanderbilt.edu}
\slugcomment{Accepted for Publication in the Astrophysical Journal Letters}
\shorttitle{Identification of the LDB and Age of Blanco~1}
\shortauthors{Cargile, James \& Jeffries}
\begin{document}
\title{Identification of the Lithium Depletion Boundary \\and Age of the Southern Open Cluster Blanco~1}
\author{P.~A. Cargile\altaffilmark{1}, D.~J. James\altaffilmark{2,3}, R.~D. Jeffries\altaffilmark{4}}
\altaffiltext{1}{Department of Physics and Astronomy, Vanderbilt University, Nashville, TN 37235, USA, \myemail}
\altaffiltext{2}{Physics and Astronomy Department, University of Hawai'i at Hilo, Hilo, HI 96720, USA}
\altaffiltext{3}{Cerro Tololo Inter-American Observatory, Casilla 603, La Serena, Chile}
\altaffiltext{4}{Astrophysics Group, Keele University, Keele, Staffordshire ST5 5BG, UK}
\begin{abstract}
We present results from a spectroscopic study of the very low mass members of the Southern open cluster Blanco~1 using the Gemini-N telescope\footnote{Based on observations with the Gemini-North Telescope, program ID: GN-2009B-Q-53}. We obtained intermediate resolution (R$\sim$4400) GMOS spectra for 15 cluster candidate members with I$\sim$14--20 mag, and employed a series of membership criteria -- proximity to the cluster's sequence in an I/I$-$K$_{s}$ color-magnitude diagram (CMD), kinematics agreeing with the cluster systemic motion, magnetic activity as a youth indicator -- to classify 10 of these objects as probable cluster members. For these objects, we searched for the presence of the \ion{Li}{1} 6708\AA\ feature to identify the lithium depletion boundary (LDB) in Blanco~1. The I/I$-$K$_{s}$ CMD shows a clear mass segregation in the Li distribution along the cluster sequence; namely, all higher mass stars are found to be Li-poor, while lower mass stars are found to be Li-rich. The division between Li-poor and Li-rich (i.e., the LDB) in Blanco~1 is found at I$=$18.78$\pm$0.24 and I$-$K$_{s} =$3.05$\pm$0.10. Using current pre-main-sequence evolutionary models we determine an LDB age of 132$\pm$24 Myr. Comparing our derived LDB age to upper-main-sequence isochrone ages for Blanco~1, as well as for other open clusters with identified LDBs, we find good chronometric consistency when using stellar evolution models that incorporate a moderate degree of convective core overshoot.
\end{abstract}
\keywords{open clusters and associations: general --- open clusters and associations: individual (Blanco~1) --- stars: evolution --- stars: fundamental parameters}
\section{Introduction}\label{sec.intro}
As pre-main-sequence (PMS) low-mass stars ($<$0.6 M$_{\odot}$) approach the zero-age main-sequence, their natal lithium content is rapidly destroyed by proton burning in regions where their interior temperature reaches above $\sim 2.5\times10^{6}$ K. Since the central temperature of a young star is a sensitive function of stellar mass and age \citep[][]{Bildsten1997,Ushomirsky1998}, determining the mass, or equivalently the luminosity, at which stars in an open cluster fully deplete their initial lithium content allows us to measure the cluster age -- the so-called lithium depletion boundary (LDB) method. The LDB stellar dating technique is model dependent; however, unlike traditional isochrone modeling of the CMD, different evolutionary LDB models yield very similar ($\pm$10\%) ages \citep{Burke2004}. It is this model-insensitivity that makes LDB ages so valuable, not only as a tool to define and constrain models of stellar evolution, but more importantly in assisting us to identify missing input physics to be included in the more widely used isochrone modeling technique.

LDB ages are inherently difficult to measure due to the necessity of procuring spectra of very faint, low-mass stars that have not yet depleted their natal lithium, typically a cluster's mid- to late-M dwarfs. Furthermore, the technique is most sensitive to young stellar ages, $\sim$10--250 Myr. These limitations have resulted in only five open clusters so far having LDB age determinations: Pleiades (126$\pm$11 Myr; \citealt{Stauffer1998,Burke2004}), $\alpha$ Persei (90$\pm$10 Myr; \citealt{Stauffer1999}), IC~2391 (50$\pm$5 Myr; \citealt{BarradoyNavascues1999,BarradoyNavascues2004}), NGC~2547 (35$\pm$4 Myr; \citealt{Jeffries2005}), and IC~4665 (27$\pm$5 Myr; \citealt{Manzi2008}). In this letter, we present initial results from our campaign to identify the LDB in the open cluster Blanco~1. 

Blanco~1 is a relatively young (50--150 Myr; \citealt{Panagi1997,Moraux2007}), nearby Southern open cluster (209 pc; \citealt{vanLeeuwen2009}) of particular astrophysical interest due to its high Galactic latitude ($b = -79^{o}$), and its comparable age to the well-studied Pleiades cluster. Considerable interest in the cluster has been driven by its reported metal-rich nature ([Fe/H]$=\ +$0.23; \citealt{Edvardsson1995}), although a more recent, self-consistent determination now makes the cluster of near-solar composition ([Fe/H]$=\ +$0.04$\pm$0.04; \citealt{Ford2005}). A combination of the cluster's systemic motion, distance below the Galactic plane ($\sim$250 pc), and estimated age (50--150 Myr) suggests that it was formed in or very near to the Galactic plane, and has subsequently moved to its current Galactic position. Its Galactic location makes it an highly attractive target for LDB study due its low level of field star contamination, making membership selection relatively straightforward \citep[e.g.,][]{Mermilliod2008}. 
\section{Target Selection, Observations, and Data Reduction}\label{sec.targetsel}
Recently, \citet[][hereafter M07]{Moraux2007} assembled an infrared photometric catalog of the very low-mass (VLM) members of Blanco~1. In order to identify brown dwarf candidates in the cluster, they observed 17 photometric candidate members having I$\sim$18.0--20.0 with low-resolution (R$\sim$1000) spectroscopy using the VLT FORS2 and KECK LRIS instruments. Using a detailed spectral analysis technique, they identified 15/17 objects as probable cluster low-mass members. Unfortunately, their spectral resolution level precluded measurement of the location of the LDB for Blanco~1. However, M07's list of probable members provides us with a robust target list for our LDB study, as the expected luminosity of the LDB in a cluster of age $\sim$100 Myr and at a distance of $\sim$200 pc, corresponds to I$\simeq$19 \citep{Burke2004}.

We obtained spectra during 2009 October 11--19 of seven probable Blanco~1 members listed in M07 with the Gemini Multi-Object Spectrograph (GMOS) in queue schedule mode on the Gemini-North telescope \citep{Hook2004}. The targets have an apparent magnitude range bracketing the expected cluster's LDB, i.e., I$=$18.3--19.7. We observed these Blanco~1 candidates in 6 separate 5$\farcm$5$\times$5$\farcm$5 fields using 1$\arcsec$ slitlets for our target stars (R$\sim$4400), employing the R831 (G5302) grating blazed at 7570\AA, with a blue-blocking OG515 (G0306) filter. This instrument setup produced spectra with a wavelength range of $\sim$5700--8000\AA\ with a resolution of 0.67 \AA\ per pixel. Exposure times were set between 65 and 110 minutes, depending on the faintest target in each field, resulting in spectra with signal-to-noise ratio (SNR) per pixel of $\sim$10 and $\sim$500 for the faintest and brightest targets, respectively. Exploiting the GMOS multi-object mode, we used a recent optical survey, performed using the SMARTS 1.0m telescope at CTIO, to select an additional 8 stars with I$_{c}\sim$13.0--17.5 found in the 6 GMOS fields that we identified as photometric candidate members from their location near the cluster sequence in an optical CMD (see James et al. in prep). Most of these stars had no previous ancillary evidence for cluster membership because their brightness falls below the faintness limit of most previous membership studies for the cluster (e.g., proper motions are currently limited to I$_{c}=$14.5; James et al. in prep). Not only do the inclusion of these stars in our program increase the total number of known Blanco~1 members, but they also help us identify the magnitude range of stars that have already depleted their lithium content, thus allowing for a more confident LDB measurement. In addition to these primary target stars, a dM6 radial-velocity (RV) standard star, GJ~905, was also observed with the identical instrument setup as our Blanco~1 objects. We reduced all of our GMOS spectra using the standard reduction routines available in the IRAF Gemini-GMOS package\footnote{IRAF, in our case through http://iraf.net, is distributed by the National Optical Astronomy Observatories, which are operated by the Association of Universities for Research in Astronomy, Inc., under cooperative agreement with the National Science Foundation.}, including bias removal, aperture extraction, and wavelength calibration. 
\section{Analysis}\label{sec.analysis}
In Table~\ref{tab1}, we list the 15 photometric cluster candidate members observed as part of our GMOS observing program. For each object, we include positions, photometric properties, RVs, H$\alpha$ equivalent widths (EW), whether we detected \ion{Li}{1} absorption at 6708\AA, and each object's membership status. Positions are taken from the 2MASS catalog \citep{Skrutskie2006} for stars with I$<$17.5 (i.e., those stars taken from our optical catalog), and for fainter stars we use the positions published in M07.
\begin{deluxetable*}{l c c c c c c c c}
\tablecolumns{9}
\tablewidth{0pc}
\tabletypesize{\footnotesize}
\tablecaption{Blanco~1 GMOS Targets \label{tab1}}
\tablehead{
  \colhead{Name\tablenotemark{a}}         &
  \colhead{R.A.\tablenotemark{b}}         & 
  \colhead{Dec\tablenotemark{b}}          & 
  \colhead{I\tablenotemark{c}}            &
  \colhead{I$-$K$_{s}$\tablenotemark{c}}  &
  \colhead{RV}                            &
  \colhead{EW(H$\alpha$)\tablenotemark{d}}  &
  \colhead{Li\tablenotemark{e}}         &
  \colhead{Member?\tablenotemark{f}}      \\
  \colhead{}                              &
  \colhead{[HH:MM:SS]}                    & 
  \colhead{[DD:MM:SS]}                    &
  \colhead{[mag]}                         &
  \colhead{[mag]}                         &
  \colhead{[km s$^{-1}$]}                  &
  \colhead{[\AA]}                       &
  \colhead{Detection}                       &
  \colhead{[Y/N]}                         }
\startdata
JCO-F18-88        & 00:01:39.86 & -30:04:38.23 & 13.345$\pm$0.003 & 1.723$\pm$0.029     &  +6$\pm$14 & +0.8$\pm$0.05 &  N & Y \\
JCO-F2-78         & 00:07:40.88 & -30:05:57.01 & 14.339$\pm$0.018 & 1.966$\pm$0.034     &  +4$\pm$7  & +3.3$\pm$0.08 &  N & Y \\
JCO-F2-48         & 00:07:56.94 & -30:04:17.08 & 14.503$\pm$0.005 & 2.101$\pm$0.031     & -68$\pm$4  & -0.3$\pm$0.08 &  N & N \\
JCO-F13-55        & 00:04:22.79 & -30:23:05.92 & 15.757$\pm$0.033 & 2.163$\pm$0.061     &  +6$\pm$9  & +7.1$\pm$0.15 &  N & Y \\
JCO-F18-133       & 00:01:33.83 & -30:06:20.02 & 15.793$\pm$0.015 & 1.858$\pm$0.064     & +33$\pm$6  & -0.4$\pm$0.08 &  N & N \\
JCO-F9-190        & 00:05:13.36 & -30:26:28.22 & 16.230$\pm$0.089 & 2.116$\pm$0.109     &  +3$\pm$5  & +4.6$\pm$0.14 &  N & Y \\
JCO-F9-229        & 00:05:22.24 & -30:27:58.99 & 16.778$\pm$0.031 & 2.204$\pm$0.093     &   0$\pm$8  & -0.4$\pm$0.08 &  N & N \\
JCO-F18-123       & 00:01:36.44 & -30:05:55.11 & 17.369$\pm$0.053 & 2.518$\pm$0.149     &  +3$\pm$5  & +0.5$\pm$0.06 &  N & N \\
CFHT-BL-16        & 00:01:28.46 & -30:06:06.56 & 18.33$\pm$\nodata  & 2.87$\pm$\nodata  &  +3$\pm$4  & +4.5$\pm$0.17 &  N & Y \\
CFHT-BL-24        & 00:07:50.59 & -30:05:09.04 & 18.54$\pm$\nodata  & 2.97$\pm$\nodata  &  +8$\pm$6  & +8.1$\pm$0.14 &  N & Y \\
CFHT-BL-28        & 23:59:55.40 & -30:02:33.54 & 18.78$\pm$\nodata  & 2.82$\pm$\nodata  & -29$\pm$4  & -0.1$\pm$0.09 &  N & N \\
CFHT-BL-38        & 00:05:13.03 & -30:27:35.65 & 19.01$\pm$\nodata  & 3.12$\pm$\nodata  &  +5$\pm$6  & +4.3$\pm$0.18 &  Y & Y \\
CFHT-BL-43        & 00:04:32.81 & -30:18:42.32 & 19.11$\pm$\nodata  & 3.15$\pm$\nodata  & +10$\pm$7  & +7.1$\pm$0.17 &  Y & Y \\
CFHT-BL-45        & 00:01:35.62 & -30:03:09.47 & 19.33$\pm$\nodata  & 3.29$\pm$\nodata  & +16$\pm$16 & +4.4$\pm$0.30 &  Y & Y \\
CFHT-BL-49        & 00:04:28.83 & -30:20:37.52 & 19.49$\pm$\nodata  & 3.58$\pm$\nodata  &  +6$\pm$14 & +6.0$\pm$0.53 &  Y & Y \\
\enddata
\tablenotetext{a}{Targets are from: JCO--SMARTS optical survey; CFHT-BL--\citet{Moraux2007}}
\tablenotetext{b}{J2000.0 Coordinates}
\tablenotetext{c}{The K$_{s}$ values are from either the 2MASS catalog for stars with I$<$17.5, or \citeauthor{Moraux2007} for I$>$17.5. For I$>$17.5, no individual photometric unceratinties were published in \citeauthor{Moraux2007}} 
\tablenotetext{d}{Positive values indicate the line is in emission.}
\tablenotetext{e}{Y--Li absorption detected at the $\sim$3-$\sigma$ level, N--Li absorption not detected above 1-$\sigma$.}
\tablenotetext{f}{Y--confirmed cluster member, N--either cluster non-member or currently undetermined.}
\end{deluxetable*}
Photometric data for our Blanco~1 targets are collected from two sources: for stars with I$<$17.5 (i.e., those indicated with a ``JCO'' prefix in Table~\ref{tab1}) we use the I$_{c}$ magnitudes from our recent photometric survey of several southern open clusters (e.g., see \citealt{Cargile2009, Cargile2010}, James et al. in prep), and K$_{s}$ photometry is from the 2MASS survey \citep{Skrutskie2006}. Uncertainties on individual photometric data points are taken from the original catalogs. For the fainter stars (I$>$17.5), we use the I and K$_{s}$ values listed in the M07 catalog. Individual uncertainties are not published in the M07 catalog; however, they do list an average photometric error of $\sim$0.04 and $\sim$0.03 mag for I and K$_{s}$, respectively. Therefore, we adopt these average uncertainties for all of the stars we include from the M07 catalog (i.e., I$>$17.5). 

RVs for each of Blanco~1 targets (see Table~\ref{tab1}) were measured by cross-correlating the GMOS spectra with the RV standard star GJ~905. Uncertainties on these RVs are relatively large ($\Delta$RV$\sim$5--15 km~s$^{-1}$), which is due to the low SNR of the target spectra and the moderate resolution of our observations.

We measure the EW of the H$\alpha$ feature for each GMOS spectra using the SPLOT task in IRAF. EW were determined using a Gaussian line-profile modeled using a linear normalization to the pseudo-continuum calculated from wavelength regions flanking the H$\alpha$ feature. Uncertainties for EW have been approximated using the formula $\Delta$EW$\sim1.5\times\sqrt{\mathrm{FWHM}\times\mathrm{p}}/\mathrm{SNR}$, where FWHM, p, and SNR are the FWHM of the Gaussian, the pixel dispersion scale in \AA, and the SNR, respectively \citep{Cayrel1988}. We include the 1-sigma errors on the H$\alpha$ EW measurements in Table~\ref{tab1}. 
\subsection{Membership Selection}\label{subsec.membership}
In order to classify our target stars as cluster members of Blanco~1, we consider three different membership criteria: (1) Photometry consistent with the cluster sequence in an I/I$-$K$_{s}$ CMD. (2) RVs must be within 1-$\sigma$ of the cluster systemic velocity \citep[+6 km s$^{-1}$][]{Mermilliod2008}. (3) H$\alpha$ line EWs must be comparable to similar-mass stars in the similar-age Pleiades cluster.

\begin{figure}[h] 
  \centering
  \includegraphics[scale=0.6,angle=-90]{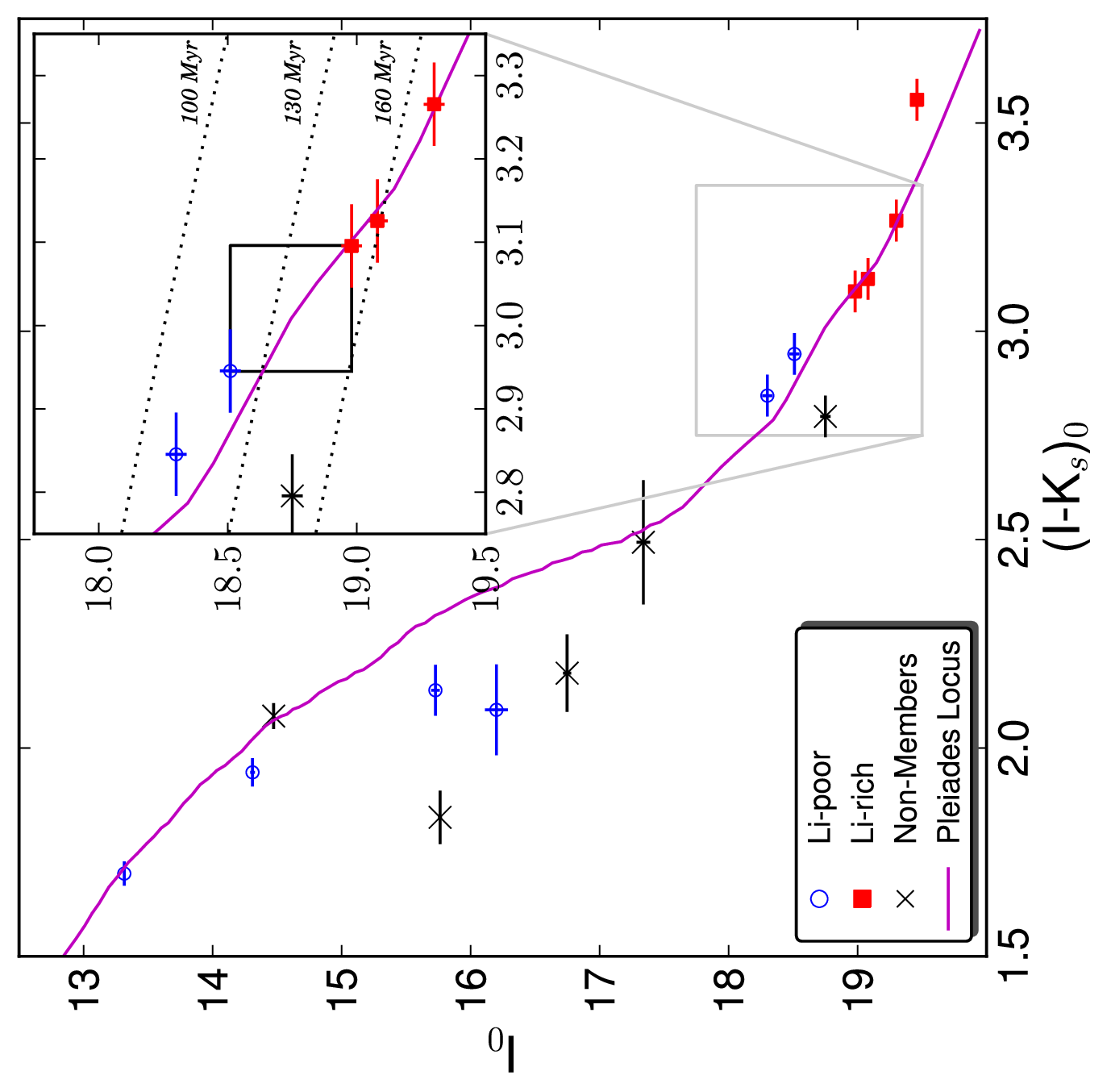} 
  \caption{
    \label{fig1}  
    Intrinsic I/I$-$K$_{s}$ CMD for 15 Blanco~1 candidates. Stars not 
    identified as single-star cluster members are identified as black crosses. 
    VLM cluster members with significant Li absorption are indicated as filled 
    red squares, and Li-poor members are shown as open blue circles. The Pleiades 
    single-star locus is plotted as a purple solid line. A zoomed-in region 
    around the LDB is plotted in the inset. The solid box shows the probable 
    region of the LDB for Blanco~1. Also plotted are BCAH98 predicted constant 
    luminosity loci (dotted lines) corresponding to the given LDB ages.}
\end{figure}

In Figure \ref{fig1}, we plot the CMD for all 15 objects observed in our GMOS program. Of these, 14 have photometry consistent, within 3-$\sigma$ of their photometric uncertainty, of the empirical, single-star I/I$-$K$_{s}$ Pleiades cluster sequence from \citet{Stauffer2007} shifted to the distance to Blanco~1 \citep[207$\pm$12 pc; ][]{vanLeeuwen2009}. We also take into account a possible 0.1 magnitude offset of the cluster sequence in order to account for natural variability in young stars. We further rejected two targets due to their RV being inconsistent with single-star membership to the cluster. In both cases, the rejected RV differed from the cluster systemic velocity by many $\sigma$; however, we note that these rejected objects might in fact be cluster members in short-period binary systems. Further RV measurements are required to confirm their binary nature. We reject an additional two stars, JCO-F9-229 and JCO-F9-123, for not having significant H$\alpha$ emission (see below). In summary, we have identified 10 high fidelity VLM members of Blanco~1, out of the 15 observed targets. 

Each membership criterion has its own level of field star contamination, therefore, we consider each individual property as necessary but not entirely sufficient on their own for cluster member rejection. By combining the three membership criteria we can be confident that stars satisfying all three properties are very likely single-star Blanco~1 members. In particular, two objects, JCO-F9-229 and JCO-F9-123, are photometric members and have RVs consistent with cluster membership, however do not have significant H$\alpha$ emission, i.e., EW(H$\alpha$)$>$+0.5\AA. These stars may be cluster members with abnormally low levels of H$\alpha$ emission, however follow-up observations are required to verify these stars as bona fide cluster members. Nevertheless, in order to be consistent in our membership selection, and to ensure the lowest probability of contamination in our membership catalog of VLM Blanco~1 stars, we presently classify these two stars as cluster non-members.
\subsection{Identification of Blanco~1 LDB}\label{subsec.LDB}
Previous studies have shown that young, VLM (spectral-type M0--M9), Li-rich stars in similar-aged open cluster have EW(Li)$\sim$1\AA\ \citep{Stauffer1998,Stauffer1999,BarradoyNavascues2004}. In fact, the \ion{Li}{1} 6708\AA\ line is expected to saturate at $\sim$0.6--0.7\AA\ according to the curves-of-growth given in \citet{Osorio2002}. Also, the saturated nature of the \ion{Li}{1} feature means its absence indicates significant Li depletion has occurred (e.g., 90\% Li-depletion only reduces the EW(Li) by a factor of $\sim$2 according to \citealt{Osorio2002}). Therefore, the presence or absence of the \ion{Li}{1} 6708\AA\ feature is an excellent indicator of PMS Li-depletion in stars. Unfortunately, the combination of a low SNR for our faintest stars and the medium resolution of GMOS makes a direct EW measurements of the Li feature in our Blanco~1 spectra very uncertain. Therefore, we employ a comparative analysis approach to detect the presence of Li in our Blanco~1 objects. 
\begin{figure}[h] 
  \centering
  \includegraphics[scale=0.425,angle=0]{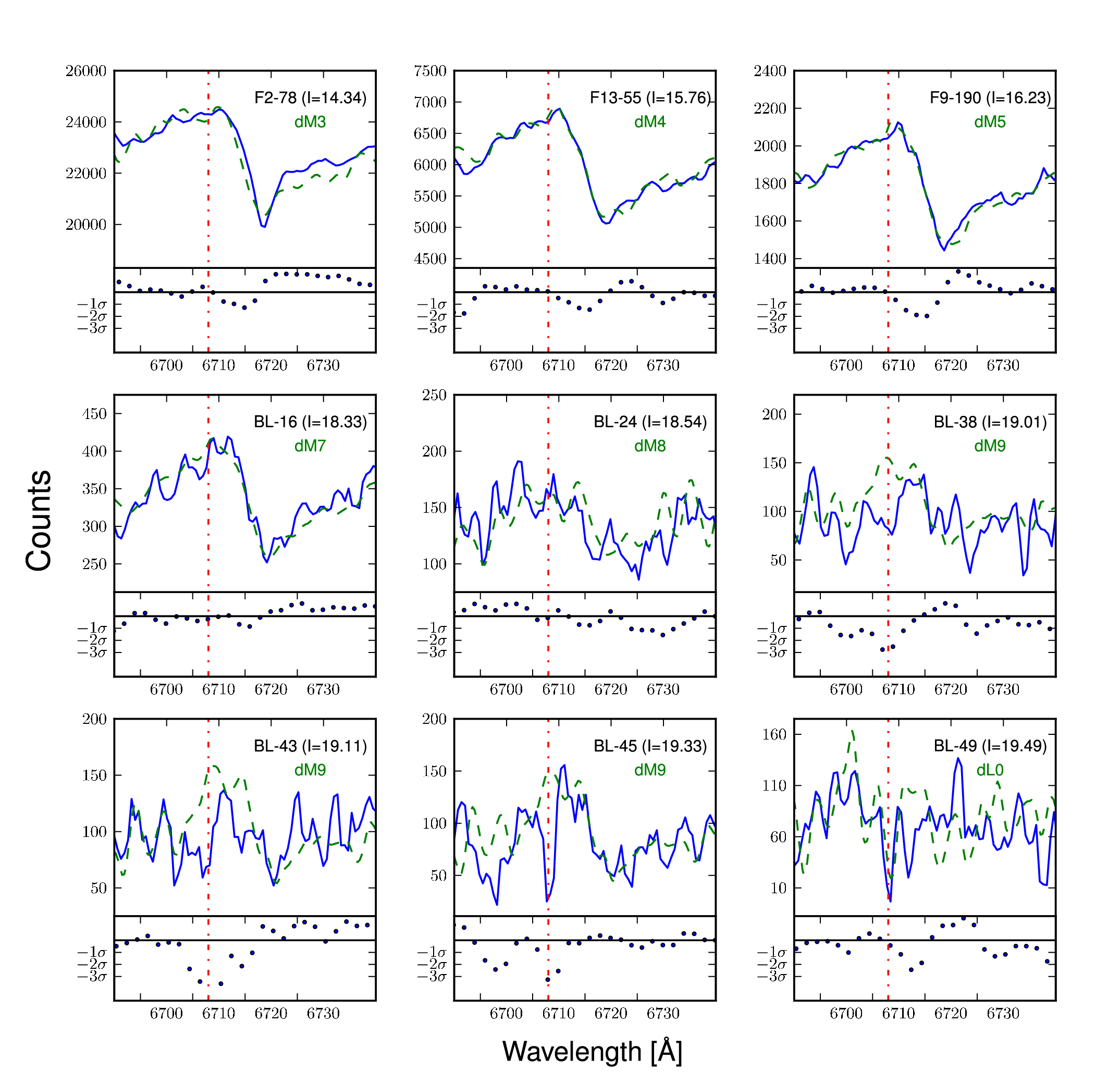} 
  \caption{
    \label{fig2}  
    GMOS spectra around the Li 6708\AA\ feature (red dotted line) 
    are plotted for 10 confirmed VLM members of Blanco~1 (blue solid line). 
    For comparison, scaled spectra of similar spectral-type SDSS templates are 
    also plotted (green dashed line). Below each VLM spectra is the moving 
    integrated residual (GMOS$-$SDSS) with $\sigma$-levels noted. Blanco~1 
    identifiers with I-band magnitudes (black), and SDSS template spectral-types 
    (green) are given at the top right of each panel. The early spectral-type star 
    JCO-F18-88 is not plotted.}
\end{figure}
Figure~\ref{fig2} shows the spectral region around the \ion{Li}{1} doublet at 6708\AA\ for the 10 Blanco~1 cluster members. For comparison, we also plot spectra of similar spectral-type stars from the M-dwarf template catalog from the Sloan Digital Sky Survey \citep[SDSS;]{Bochanski2007}. These templates were produced by averaging over 4000 SDSS stellar spectra for spectral-types dM0--dL0. Due to the nature of the SDSS, the majority of the combined spectra used for these templates are field M-dwarfs, and therefore are expected to be old enough ($>$0.5 Gyr) to have destroyed their initial lithium -- except for the dL0 template which is expected to be below the mass necessary to reach a central Li-depleting temperature. We matched the SDSS spectra by determining the closest match to the major absorption features and overall shape of the pseudo-continuum of the GMOS spectra.

Below each Blanco~1 spectrum is a subplot of the moving integrated residual for the GMOS and SDSS spectra. These data points are derived by integrating over the GMOS$-$SDSS residual in a moving 5\AA\ window, with a moving step-size of 2\AA. By integrating over a moving window, we average out the small difference between the GMOS and SDSS spectra (e.g., different spectral resolutions), and provide a better description of the significant deviation between the GMOS and SDSS spectra. The $\sigma$-levels are derived from the r.m.s. of the residual outside of a 10\AA\ window centered on the Li 6708\AA\ feature. We note that CFHT-BL-49 is best matched with the dL0 template which still retains its natal Li content; therefore, as expected the residual does not show a significant difference at 6708\AA.

A clear trend is present when looking at the distribution of Li-absorption in VLM Blanco~1 stars. Namely, brighter (higher mass) Blanco~1 members are ubiquitously Li-poor; any observed Li absorption at 6708\AA\ is not significant beyond 1-$\sigma$. This suggests these stars have already gone through their PMS Li burning stage. Fainter (lower mass) Blanco~1 VLM stars are all observed to be Li-rich with Li absorption detections at or above $\sim$3-$\sigma$. The transition from Li-poor to Li-rich is bracketed by CFHT-BL-24 (I$=$18.54$\pm$0.04, I$-$K$_{s}=$2.97$\pm$0.05) and CFHT-BL-38 (I$=$19.01$\pm$0.04, I$-$K$_{s}=$3.12$\pm$0.05), defining the color/magnitude range in which the LDB is located. We contend that the LDB for Blanco~1 is located at I$=$18.78$\pm$0.24 and I$-$K$_{s} =$3.05$\pm$0.10, as represented by the box in the inset of Figure~\ref{fig1}.
\subsection{LDB Age of Blanco~1}\label{subsec.Age}
Having identified the color/magnitude of Blanco~1's LDB, we can now determine its age using the predicted Li-depletion rates from current PMS evolutionary models, specifically the models from \citet[][hereafter BCAH98]{Chabrier1997,Baraffe1998}. Here, we assume the LDB is located at the luminosity where 99\% of the Li is predicted to be depleted in a star. The high sensitivity of the rate of Li-depletion to stellar mass (or luminosity) means assuming the LDB is instead located at 90\% depletion would change the measured LDB age by only $\pm$1 Myr \citep[see][]{Jeffries2005}.
\begin{deluxetable}{l c c}
\tablecolumns{3}
\tablewidth{0pc}
\tablecaption{LDB Parameters for Blanco~1 \label{tab2}}
\tablehead{
  \colhead{Parameter}    &
  \colhead{BCAH98 \tablenotemark{a}} &
  \colhead{BCAH98 \tablenotemark{a}} \\
  \colhead{} &
  \colhead{$+$L96} &
  \colhead{$+$DUSTY}
}
\startdata
M$_{I}$                & \multicolumn{2}{c}{12.17$\pm$0.24}         \\
(I$-$K$_{s}$)$_{0}$    & \multicolumn{2}{c}{3.02$\pm$0.08}           \\
M$_{bol}$                &   11.99$\pm$0.30       & 12.01$\pm$0.23   \\
T$_{eff}$ [K]            &   2780$\pm$80          & 2810$\pm$60      \\
Log(L) [L/L$_{\odot}$]   &   -2.94$\pm$0.12       & -2.90$\pm$0.09   \\
Mass [M$_{\odot}$]       &   0.074$\pm$0.05       & 0.075$\pm$0.05   \\
LDB Age [Myr]           &   132$\pm$24           & 124$\pm$18        \\
\enddata
\tablenotetext{a}{BCAH98$+$L96 -- BC from \citet{Leggett1996}; BCAH98$+$DUSTY -- theoretical bolometric corrections derived from DUSTY model atmospheres.}
\end{deluxetable}
We calculate M$_{I}=12.17$ and (I$-$K$_{s}$)$_{0}=$3.02 for the LDB of Blanco~1 using an intrinsic distance modulus from HIPPARCOS \citep[6.58$\pm$0.12,][]{vanLeeuwen2009}, and adopt E(I$-$K$_{s}$)$=$0.02 and A$_{I}=$0.03. Converting intrinsic LDB colors/magnitudes to the bolometric luminosity necessary for comparison to predictions from PMS models, we use two different bolometric corrections (BC). First, we use an empirical I/I$-$K$_{s}$ to BC relationship given by \citet{Leggett1996}. Second, we employ the BC and color-T$_{eff}$ relationships calculated by BCAH98 using DUSTY non-gray model atmospheres \citep{Baraffe2002}. In order to accurately compare the BCAH98 models to our observations, we convert their predicted CIT K photometry to the 2MASS K$_{s}$ using the relationship in \citet{Carpenter2001a}. Listed in Table~\ref{tab2} are the physical parameters predicted by BCAH98 models for our measured M$_{I}$/I$-$K$_{s}$ location of Blanco~1's LDB with its respective uncertainty. We calculate a LDB age of 132$\pm$24 Myr using the empirical BC, and a slightly younger age, 124$\pm$18 Myr, using the BC from BCAH98$+$DUSTY; however within their errors, both ages are in agreement.

The uncertainties we place on the LDB ages for Blanco~1 are derived solely from observational errors in the LDB's M$_{I}$ and intrinsic I$-$K$_{s}$ values: the error in the distance modulus (0.1 mag), the photometric uncertainty in the objects ($\sigma_{I}=$0.04 mag, $\sigma_{I-K_{s}}=$0.05 mag), and, most significantly, the error in the LDB location in the I/I$-$K$_{s}$ CMD. The precision of the LDB's location is defined by the color/magnitude difference between CFHT-BL-24 and CFHT-BL-38, as illustrated by the box in the inset of Figure \ref{fig1}; $\sigma_{I,LDB}=$0.24 mag, $\sigma_{I-K_{s},LDB}=$0.10 mag.
\section{Discussion and Implications}\label{sec.discuss}
Upper-main-sequence (UMS) isochrone ages derived using high-mass stellar evolutionary models that do not include convective-core overshooting are seen to be systematically $\sim$1.5 times younger than LDB ages \citep[e.g.,][]{BarradoyNavascues2004}. Convective-core overshoot has the effect of mixing more hydrogen into stellar cores, hence prolonging stellar main-sequence lifetimes. Therefore, increasing the amount of core overshoot can thereby bring into agreement measured UMS and LDB ages. 
\begin{figure}[h] 
  \centering
  \includegraphics[scale=0.45,angle=-90]{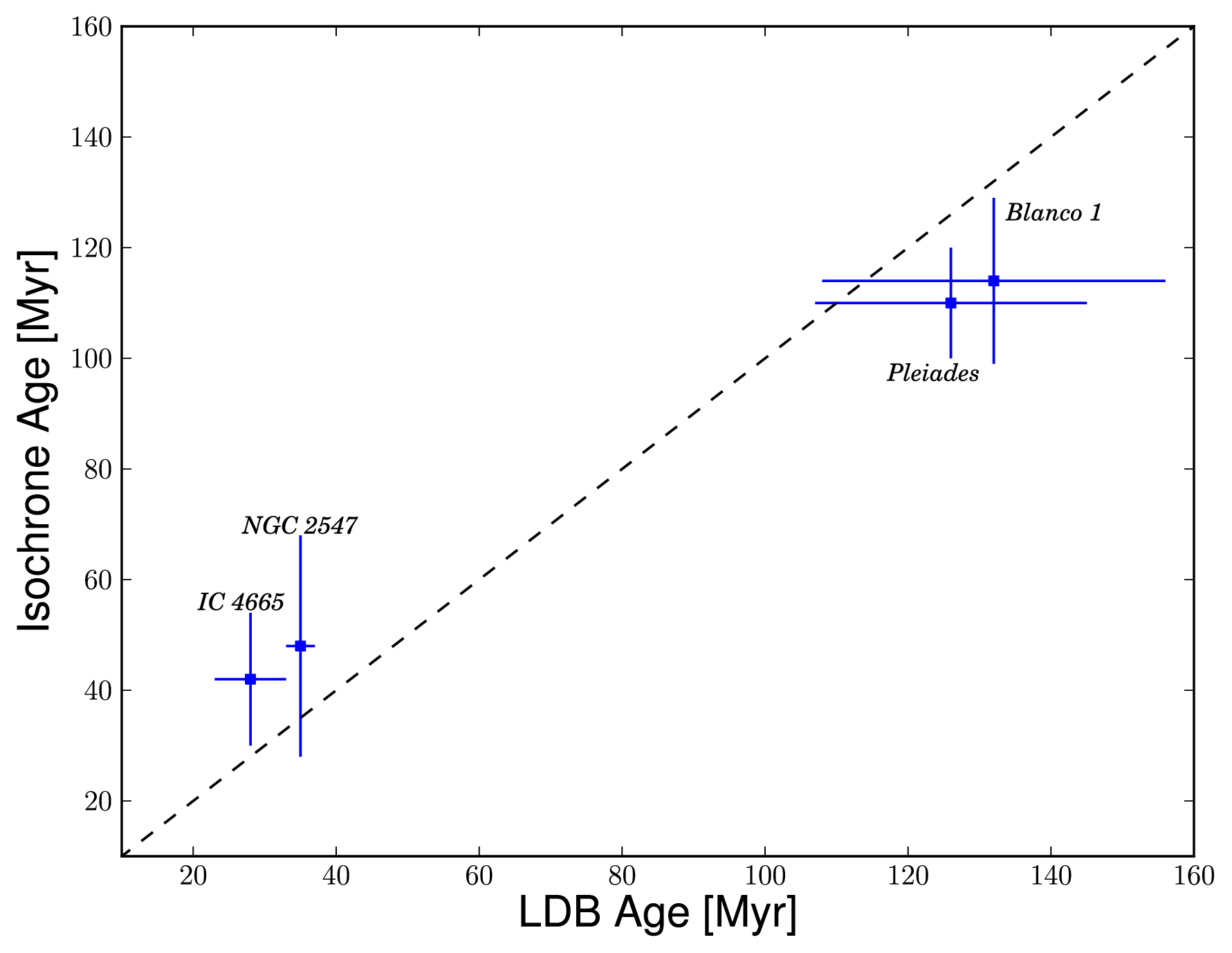} 
  \caption{
    \label{fig3}  
    LDB versus UMS age for four clusters with UMS ages found using the 
    $\tau^{2}$ fitting technique and the Geneva stellar evolutionary models. 
    The black dashed line represents equality between the two age determination 
    methods and is {\it not} a fit to the data.}
\end{figure}
Of the six open clusters with identified LDBs, including our new identification in Blanco~1, four (IC~4665, \citep{Cargile2010}; NGC~2547, \citep{Naylor2006}; Pleiades, \citep{Naylor2009}; Blanco~1, James et al in prep) also have UMS ages measured using the recently developed $\tau^{2}$ isochrone fitting technique \citep[for details, see][]{Naylor2006,Naylor2009}, and with the same high-mass stellar evolutionary models \citep[Geneva models from][]{Schaller1992,Lejeune2001} that use moderate amounts of core overshooting as determined from empirical constraints. In Figure~\ref{fig3}, we plot the LDB and UMS ages for these four open clusters. We find that all four clusters have UMS and LDB ages that are consistent within their uncertainties.

Although inclusion of a moderate degree of core overshoot is mostly likely necessary, the chronometric inconsistencies seen between LDB and isochronal ages {\em must} also be influenced by inconsistencies in PMS models. For example, if we use M$_{K_{s}}$ at the LDB instead of M$_{I}$, we find an age of 150$\pm$15 Myr using the theoretical BCs of BCAH98. This suggests a possible range of at least $\sim$20 Myr, or $\sim$15\% uncertainty at an age of 130 Myr, when comparing LDB ages derived using different colors/magnitudes predicted by the DUSTY model atmospheres. In fact, BCAH98$+$DUSTY stellar evolutionary models are known to have difficulty reproducing colors/magnitudes of young, low-mass stars \citep[see ][]{Stauffer2007}, which is attributed to missing opacities at blue wavelengths \citep{Baraffe1998}. By comparison, if we derive the LDB age of Blanco~1 using M$_{K}$ and empirically-calibrated BCs, we derive an age of 134$\pm$26 Myr -- a difference of only 2 Myr from our LDB age calculated using M$_{I}$. This agreement suggests that we can account for some of the systematic inconsistencies in theoretical photometry by deriving LDB ages using empirically constrained BC-color relationships \citep[e.g.,][]{Leggett1996}. However, both theoretical and empirical BC relationships may still contain additional systematic uncertainteis due to missing physics that is not being taken into account in the PMS evolutionary models, e.g., the influence of high magnetic activity on low-mass PMS stellar evolution \citep[see][]{Stauffer2003,Chabrier2007,Yee2010}. 
\acknowledgments
P.A.C. and D.J.J. acknowledge the support from the National Science Foundation Career Grant AST-0349075 (Principal Investigator: K.~G. Stassun). We gratefully acknowledge the staff at the Cerro Tololo Observatories and those of the SMARTS Consortium. Our research is based on observations obtained at the Gemini Observatory, which is operated by the Association of Universities for Research in Astronomy, Inc., under a cooperative agreement with the NSF on behalf of the Gemini partnership: the National Science Foundation (United States), the Science and Technology Facilities Council (United Kingdom), the National Research Council (Canada), CONICYT (Chile), the Australian Research Council (Australia), Minist\'{e}rio da Ci\^{e}ncia e Tecnologia (Brazil) and Ministerio de Ciencia, Tecnolog\'{i}a e Innovaci\'{o}n Productiva (Argentina). 
\end{document}